\pdfoutput=1

\documentclass[conference]{IEEEtran}

\makeatletter                   
\def\mdseries@tt{m}             
\makeatother 

\usepackage{listings}
\usepackage{graphicx}
\usepackage{pgfplots}
\usepackage{lscape}
\usepackage{tabularx}
\usepackage{textcomp}
\usepackage{multirow}
\usepackage[draft,inline=true,margin=false]{fixme}
\usepackage[capitalize]{cleveref}
\usepackage{xspace}
\usepackage{csquotes}
\usepackage{makecell}
\usepackage{tikz}
\usepackage[caption=false]{subfig}
\usepackage{url}
\usepackage{caption}

\usepackage{amssymb}
\usepackage{pifont}
\newcommand{\cmark}{\ding{51}}%
\newcommand{\xmark}{\ding{55}}%

\fxusetheme{color}
\usepackage{listings}
\usepackage{color}
\definecolor{lightgray}{rgb}{.9,.9,.9}
\definecolor{darkgray}{rgb}{.4,.4,.4}
\definecolor{purple}{rgb}{0.65, 0.12, 0.82}

\usepackage{algorithm,algpseudocode}
\let\oldReturn\Return
\renewcommand{\Return}{\State\oldReturn}
\usepackage[newfloat, draft=true]{minted}
\setminted[]{breaklines=true,
  fontsize=\scriptsize}

\lstdefinelanguage{JavaScript}{
	keywords={typeof, new, true, false, catch, function, return, null, catch, switch, var, if, in, while, do, else, case, break},
	keywordstyle=\color{blue}\bfseries,
	ndkeywords={class, export, boolean, throw, implements, import, this},
	ndkeywordstyle=\color{darkgray}\bfseries,
	identifierstyle=\color{black},
	sensitive=false,
	comment=[l]{//},
	morecomment=[s]{/*}{*/},
	commentstyle=\color{purple}\ttfamily,
	stringstyle=\color{red}\ttfamily,
	morestring=[b]',
	morestring=[b]",
	escapechar=|
}

\renewcommand{\paragraph}[1]{\vspace{0.5em}\noindent\textbf{#1}.}

\usepackage{shortcuts}

\begin{document}

\author{
    \IEEEauthorblockN{Behnaz Hassanshahi, Hyunjun Lee, Paddy Krishnan, J{\"o}rn G{\"u}y Su{\ss}}
    \IEEEauthorblockA{Oracle Labs, Brisbane, Australia
    \\\{\textit{firstname.lastname}\}@oracle.com}
}

\title{Gelato: Feedback-driven and Guided Security Analysis of Client-side Web Applications}

\maketitle

\begin{abstract}
Even though a lot of effort has been invested in analyzing
client-side web applications during the past decade, the existing
tools often fail to deal with the complexity of
modern JavaScript applications.
However, from an attacker point of view,
the client side of such web applications can reveal invaluable information
about the server side. In this paper, first we study the existing tools and enumerate the most crucial
features a security-aware client-side analysis
should be supporting.
Next, we propose \tool to detect vulnerabilities in
modern client-side JavaScript applications that are built upon complex libraries and frameworks.
In particular, we take the first step in
closing the gap between state-aware crawling and client-side
security analysis by proposing a feedback-driven
security-aware guided crawler that is able to analyze complex frameworks
automatically, and increase the coverage of security-sensitive
parts of the program efficiently. Moreover, we propose a new lightweight client-side taint analysis
that outperforms the start-of-the-art tools, requires no modification to browsers, and reports non-trivial
taint flows on modern JavaScript applications.
\end{abstract}

\section{Introduction}
\label{sec:intro}

The client-server design of web applications
allows developers to perform the logic
and computation of their applications on both the client and server sides.
The powerful and rich features of modern browsers have enticed
developers to use languages, such as JavaScript, HTML and CSS,
to implement complex and highly interactive user interfaces on the
client side. All the client-side source-code that runs in the
browsers is available to everyone, including attackers.
While attackers have been actively exploiting client-side vulnerabilities,
such as DOM-based XSS~\cite{owasp_domxss} for almost a decade now, the
increase in complexity of JavaScript applications and frameworks
has rendered the existing tools ineffective to prevent them.
Moreover, from an attacker point of view,
the client side of web applications can reveal invaluable information
about the server side, such as REST end points, validation routines,
and database queries.

In this paper, we propose \tool to detect vulnerabilities in
modern client-side JavaScript applications,
which are often built upon complex libraries and frameworks. More specifically,
we address the practical challenges necessary to deal with such
web applications. For this purpose,
first we enumerate the most crucial
features a security-aware client-side analysis
should be supporting, and report on the status of state-of-the-art
dynamic analysis tools accordingly.
We limit our study to dynamic analysis tools because
they suit the dynamic nature of JavaScript applications.

Even though a lot of effort has been invested in analyzing
client-side web applications during the past decade, the existing
tools often fail to deal with the complexity of
modern JavaScript applications. To understand why,
we provide an overview of the most essential
features necessary to analyze a client-side application to detect security
issues. Table~\ref{tbl:comparison}
summarizes the features supported by the existing tools and compares
them against \tool.

From this table, we can make several interesting observations:
(1) all of the tools except for \tool
miss at least six features
necessary to analyze modern real-world JavaScript applications;
(2) existing tools either
focus on improving the crawling technology or security analysis,
and there is no tool that focuses on both aspects, while one is
not independent of the other;
(3) most of the crawlers that focus on improving the coverage
(lines of JavaScript code or number of discovered hyperlinks)
are not guided towards specific locations, which can be essential
for security analysis in practice; and
(4) \tool is the only tool that directly addresses complex libraries and frameworks,
reducing the need to have manually crafted models.

Having support for all of the features listed in Table~\ref{tbl:comparison}
to perform a client-side security analysis is challenging.
In this paper, we take the first step in bringing
together a state-aware crawler and a client-side
security analysis, and in closing the gap between them.
One of the challenges that state-aware crawlers face is that
the search space they need to explore
(number of paths in the state graph) can grow exponentially.
Therefore, traversing the whole search space can result in
poor performance. In practice, we have found the efficiency
of crawlers to be one of the most essential factors
for them to be used for testing.
However, it is possible to devise an algorithm
to cover specific paths of interest efficiently,
without having to traverse the whole search space.
By guiding \tool towards specific
targets, we improve the performance while achieving
acceptable coverage.
Furthermore, \tool tries to incorporate
most of the features in Table~\ref{tbl:comparison}
that are crucial for a security analysis to detect
vulnerabilities effectively in practice.

To analyze modern JavaScript applications for security vulnerabilities, crawlers play
a substantial role by providing inputs to the program. However, the state-of-the-art security analysis
tools~\cite{lekies,dexterjs} either don't have a crawler or provide
an insufficient support. Hence, while they can report vulnerabilities
in the first page of applications for Alexa top~\cite{alexatop} websites, they cannot
report vulnerabilities in the other parts of the applications as shown in Sec.~\ref{sec:evaluation}.
During the past few years, many crawling
techniques have been proposed to explore
the search space on the client side.
\begin{landscape}
  \begin{table}
    \centering
    \small
    {
	\begin{tabular}{|l|c|c|c|c|c|c|c|c|c|c|c|} \hline
		Feature        & \tool      & Crawljax~\cite{crawljax} & j\"{A}k~\cite{jaak} & Feedex~\cite{feedex} & WATEG~\cite{wateg} & Artemis+SID~\cite{artemisSID} & Artform & DexterJS~\cite{dexterjs} & Kudzu~\cite{kudzu} & CTT~\cite{lekies} \\ \hline
		F1             & \cmark & \cmark & \cmark & \cmark & \cmark & \cmark & \cmark & \xmark & \xmark & \xmark \\ \hline
		F2             & \xmark & \xmark & \xmark & \xmark & \xmark & \cmark & \xmark & \xmark & \xmark & \xmark \\ \hline
                F3             & \cmark & \xmark & \xmark & \xmark & \xmark & \cmark & \cmark & \cmark & \cmark & \xmark \\ \hline
                F4             & \cmark & \cmark & \cmark & \cmark & \cmark & \cmark & \cmark & \cmark & \cmark & \xmark \\ \hline                
		F5             & \cmark & \xmark & \cmark & \xmark & \xmark & \xmark & \xmark & \xmark & \xmark & \xmark \\ \hline
		F6             & \cmark & \cmark & \cmark & \cmark & \cmark & \cmark & \cmark & \xmark & \xmark & \xmark \\ \hline
		F7             & \cmark & \xmark & \cmark & \cmark & \cmark & \cmark & \cmark & \xmark & \xmark & \xmark \\ \hline
                F8             & \cmark & \xmark & \xmark & \xmark & \xmark & \xmark & \cmark & \xmark & \xmark & \xmark \\ \hline
                F9             & \cmark & \xmark & \xmark & \xmark & \xmark & \xmark & \cmark & \xmark & \cmark & \xmark \\ \hline
                F10            & \xmark & \xmark & \xmark & \xmark & \xmark & \xmark & \xmark & \xmark & \xmark & \xmark \\ \hline
                F11            & \cmark & \xmark & \cmark & \xmark & \xmark & \xmark & \xmark & \cmark & \cmark & \cmark \\ \hline
                F12            & \cmark & \xmark & \xmark & \xmark & \xmark & \xmark & \xmark & \xmark & \xmark & \xmark \\ \hline
                F13            & \xmark & \xmark & \xmark & \cmark & \cmark & \xmark & \cmark & \xmark & \xmark & \xmark \\ \hline
                F14            & \cmark & \cmark & \cmark & \cmark & \xmark & \cmark & \cmark & \xmark & \cmark & \xmark \\ \hline
                F15            & \cmark & \xmark & \xmark & \xmark & \xmark & \xmark & \xmark & \xmark & \xmark & \xmark \\ \hline
                
        \end{tabular}       
    }
    \captionsetup{singlelinecheck=off,font=small,justification=centering}
  \caption[]{Comparing features supported by existing client-side analysis tools. Below, we provide a short description for each of these features:
    \begin{itemize} \item[] \textbf{F1: Client-state-aware} refers to crawlers that
interact with the client-side user interface and explore
its different possible states at runtime.
\item[] \textbf{F2: Server-state-aware}
  refers to crawlers that aim to trigger different states at the server side.
  
\item[] \textbf{F3 \& F4: Data input \& event sequence generation}
  shows whether the analyzer generates data inputs and event
  sequences, respectively.

\item[] \textbf{F5: Dynamic event handlers registration}
  shows whether the analyzer supports event handlers that are registered dynamically.   

\item[] \textbf{F6: Static link extraction} refers to the crawlers that can statically
  find hyperlinks embedded into HTML pages.
  
\item[] \textbf{F7: Prioritization} shows whether a crawler
 can prioritize the triggering of certain events to achieve a specific goal, such as
improving coverage.

\item[] \textbf{F8: Filling forms} is required in some JavaScript
  applications, where users enter data via forms. This feature
  shows whether the crawler is able to fill such forms.

\item[] \textbf{F9: Bypassing guards in the code} is required to
  explore applications deeply and reach specific locations
  in the program. String constraint solving is a popular technique
  used to generate such inputs.

\item[] \textbf{F10: Handling nondeterminism} is required
  when the state-aware crawler
  cannot go to a previously visited state, or
  gets trapped in one state and cannot make progress due to usage of
time stamps, randomization and being dependent on the existence of
certain behaviors that change over time.

\item[] \textbf{F11: Security analysis} refers to having support for
client-side vulnerability detection techniques, such as
DOM-based XSS detection.

\item[] \textbf{F12: Guiding towards sinks} is a feature that allows
a crawler to drive a security analysis more efficiently and
effectively.

\item[] \textbf{F13: Triggering specific functionalities}
  through a sequence of events is required by some client-side applications
  that provide complex functionalities.

\item[] \textbf{F14: Improving coverage} is often the goal of all the
crawlers. Coverage usually is measured in terms of the number of
detected hyperlinks or the amount of executed JavaScript code.

\item[] \textbf{F15: Support for modern libraries and frameworks}
  is required for most of the modern JavaScript
  applications that are built on top
of complex frameworks, such as React~\cite{react.js}, Knockout.js~\cite{knockout}.

\end{itemize}
  }
   \label{tbl:comparison}
  \end{table}
\end{landscape}

Given the large search space,
our key insight is that a crawler needs to be guided to accelerate
a target dynamic analysis to solve a problem. We propose a hybrid analysis
of JavaScript applications that is feedback-driven to direct crawlers towards reaching
program locations of interest, such as DOM-based XSS sinks and REST calls.

Once the crawler identifies endpoints and drives the execution to reach
them, a security analysis runs to report security vulnerabilities.
In this work, we focus on DOM-based XSS and reflected XSS vulnerabilities.
The DOM-based XSS analysis, in particular, requires a practical dynamic taint analysis to work
on real-world applications. Being able to analyze
modern real-world applications as our main objective,
we propose a novel staged taint inference analysis to detect DOM-based XSS
vulnerabilities. Compared to the state-of-the-art
dynamic taint analysis tools~\cite{affogato, dexterjs, jalangi1, linvail, lekies, domsday}, our solution
has better recall and is less intrusive, which makes our analysis
less likely to break the semantics of the applications.

The rest of the paper is organized as follows:
Sec.~\ref{sec:relwork} summarizes the closely related works and how they compare
against \tool. Sec~\ref{sec:design} describes the design of our feedback-driven and guided crawler, and the
input value generation to bypass guards in the program. Sec.~\ref{sec:taintinference} explains the
staged taint inference analysis used to detect DOM-based XSS vulnerabilities. Sec.~\ref{sec:implementation}
provides details on the implementation of \tool and Sec.~\ref{sec:evaluation} explains the experimental setup,
evaluation results on various benchmarks and applications, and comparison against state-of-the-art tools.
Finally, Sec.~\ref{sec:conclusion} concludes the paper. In summary, we make the following contributions:
\begin{itemize}
\item{a new crawler that can be guided towards program locations of interest using a call graph.}
\item{a feedback-driven analysis that enables our guided crawler to support modern client-side JavaScript libraries and frameworks instead of using manually crafted models.}
\item{a novel staged taint inference analysis that detects potential DOM-based XSS vulnerabilities with high accuracy.}
\item{an input generator that supports both event and data value generation to increase the coverage of security analyses.}
\end{itemize}

\section{Related Work}
\label{sec:relwork}

Client-side web application analysis has a large
body of literature, and over the past decade, many crawling
and security analysis techniques have been developed.
In this section, we compare \tool against the related works
in terms of the crawling technology and taint analysis.

\subsection{Client-side web application crawling}

Crawljax \cite{crawljax} is a state-aware crawler
that explores AJAX-based client-side applications
using dynamic analysis. It computes the edit distance
of the string representation of DOM trees
to compare states, and performs both
depth-first and breadth-first search strategies.
Compared to Crawljax, our approach is targeted and
aims to increase coverage for a set of program
locations of interest. We combine static and dynamic
analysis and also generate data values as inputs to guide
the runtime execution.

j\"{A}k \cite{jaak} is 
designed to analyze modern web applications to find
server-side security vulnerabilities.
The main goal of this work is to increase
code coverage and trigger all interaction points
in the client-side program to be able to find more vulnerabilities on the server side.
For this purpose,
it uses dynamic analysis of the client-side JavaScript program to detect
dynamically generated URLs, the registration of events, etc.
The dynamic analysis is combined with crawling to interact with
the application and infer
a navigational model.

Similar to j\"{A}k, we perform dynamic analysis
to collect runtime values, traces and to capture
events that are difficult to detect statically.
Additionally, we use static analysis to guide
the crawler to specific locations in the program
to avoid the state explosion problem. Moreover,
we integrate input value generation to our
crawling technique to increase the coverage
of relevant parts of the program.

FEEDEX~\cite{feedex} uses a state-aware feedback-directed crawling technique to derive
a test model for client-side web applications. The main focus of FEEDEX
is to reduce the test model size and enhance coverage in three aspects of
a test model: functionality, navigation and page structure.
As FEEDEX crawls the web application, the coverage of
these three aspects is fed back to the tool to prioritize
next states and events. Compared to FEEDEX, our technique
focuses on guiding the crawler towards
specific locations in the program. We guide the analysis by combining
static analysis, link extraction, state-aware crawling and input value generation.
To analyze modern web applications that heavily use complex libraries
for which static analysis is difficult,
our feedback-directed analysis refines the
static analysis results using the runtime execution.
This novel design allows us to use the crawler
to improve coverage of security analyses, such as DOM-based XSS detection and
black-box REST fuzzing.

WATEG~\cite{wateg} takes a rule-directed test-case generation approach
that focuses on increasing the coverage
of business rules (functionalities) provided as specifications.
As the program executes, WATEG checks
whether the invariants  derived from
business rules are not violated.
In this work, the crawler is directed towards pre-determined
functionalities (business rules)
using a two-phase approach: 
(1) uses coarse state comparison
to create STD (abstract state transition diagram) and
to determine the portions of the state space that are relevant;
(2) uses a more fine-grained state comparison for the relevant parts.
The abstract paths of STD are used in the second phase as starting
points and refined to traversable paths that lead to triggering the
pre-determined functionalities.

Even though our technique also
guides the execution of the program, its goal is different. Unlike WATEG,
our technique guides the execution towards specific locations
in the JavaScript code or events and not to cover pre-defined business
rules. Therefore, we  analyze the JavaScript code
and use program analysis techniques to prioritize
the event and state prioritization. 

In what follows, we provide an overview of other
related works (not closely related) in the literature
and compare against our technique.

\subsubsection{Scoped and guided crawlers}
Scoped crawling~\cite{survey2010} of client-side web applications proposes
strategies to limit the scope of the exploration
based on textual content, such as topic, geography,
language, etc. However, our approach uses program analysis
techniques to guide the execution towards specific program locations.

Guided crawling aims to guide the exploration
to achieve a particular goal, such as increasing
code, functionality or navigation coverage~\cite{artemis,feedex,wateg}. Our crawling technique is also guided because it
uses prioritization strategies to guide the execution.
However, it aims to increase coverage
of specific program locations to drive a target analysis.

\cite{efficient1} guides the exploration of the application, aiming to discover
as many states as possible in a given amount of time. Compared
to~\cite{artemis,feedex,wateg}, which focus on increasing
the diversity of crawled pages, this work mainly focuses
on increasing efficiency for an anticipated model. Our crawling technique
also improves efficiency for reaching program locations
of interest without requiring a crawling model, and uses different
analysis techniques to make it targeted.

Recently, guided fuzzing techniques have been
proposed for C/C++ programs~\cite{aflgo,vuzzer}.
Similar to our approach, AFLGo~\cite{aflgo} uses call graph
distance from target locations in the program as a metric to prioritize
input generation. However, our approach goes
one step further and also allows on-the-fly refinement
of the statically constructed call graph
to add missing and remove false positive edges in the call graph.
Moreover, we analyze JavaScript applications
that have a highly dynamic nature and require
support for both events and input values.

\subsubsection{Traditional crawlers}
Many of the traditional web application crawlers available in the industry are not state-aware, and
rely only on static link extraction over the HTML pages~\cite{arachni, zap, w3af}.
To be practically
useful for vulnerability detection, security analysts need to
manually interact with the application to take
the browser to the desired states. However, our technique is
fully automatic, and combines static and dynamic analysis
of JavaScript code to guide the crawler towards target (security-sensitive)
locations in the program.

\subsubsection{Data input generation}
In addition to event-based inputs (e.g., clicking),
client-side web applications also accept input value
in URLs and form elements such as input fields.
Therefore, to explore a client-side application deeply enough,
input value generation is required. For input fields, existing
approaches mostly provide random data if
no custom data is available~\cite{atusa}.
More heavyweight analyses such as symbolic execution
have also been proposed~\cite{kudzu,artform}, which are
known to be prone to scalability issues.
In contrast, we propose a lightweight data input
generation technique based on taint inference that
starts with random or custom data, but generates
new inputs to bypass validation routines and reach target locations (sinks)
in the program.

Autogram is a recent work that explores input generation by mining grammars~\cite{autogram}.
It uses dynamic taint tracking to trace the data flow of each
input character for a set of sample inputs. By grouping
input fragments that are handled by the same functions, it produces
a context-free grammar that can be combined with fuzzers
to generate inputs. In contrast, our input generation
assumes that the input grammar is known (e.g., URL) and
tries to bypass the validation routines using taint inference.

\subsection{Taint Analysis}

In this section, we compare our approach against existing
taint analysis techniques
used for finding vulnerabilities in web applications.

Several static analysis techniques have also been proposed to analyze
JavaScript applications~\cite{TAJS, WALA, SAFE}. 
However, the lack of 
static predictability and presence of dynamic typing in JavaScript as well as
the asynchronous and event-based nature of web applications can be highly problematic for 
determining taint flows using static analysis techniques.

Dynamic analysis of JavaScript programs 
requires instrumentation.
Two types of instrumentation techniques are used for dynamic analysis 
of web applications: engine-level instrumentation \cite{lekies, vogt,domsday}, and code-rewriting 
\cite{jalangi1,dexterjs, jstaint,affogato}.
Engine-level instrumentation involves adding hooks to the JavaScript engine.
While this design can have performance 
benefits from being compiled into the engine itself, it is not portable across different engines and requires considerable effort to maintain~\cite{lekies}.

On the other hand, source code-level instrumentation~\cite{jalangi2, linvail} involves replacing and/or
appending code to the existing program's source-code so that the runtime behavior can be analyzed without 
affecting the original behavior. This method has drawbacks in 
performance, but is often easier to write and test, and can be engine-agnostic.

Dynamic taint analysis involves tracking taint labels in a program during 
its execution. DexterJS~\cite{dexterjs} carries
out character-level taint tracking using code-rewriting to discover 
potentially vulnerable taint flows. This approach involves tracking each character 
originating from a taint source individually. Even a single character reaching 
a sink can be detected, provided that it is propagated from the tainted value.
DexterJS attaches taint labels to primitive values by wrapping (boxing) them.
Because built-ins, browser APIs and DOM functions cannot be instrumented,
DexterJS requires hard-coded models.
However, coming up with models
that capture all runtime behavior is very challenging. In fact, our
experiments with DexterJS~\cite{dexterjs} show that incomplete models for
built-ins result in missing valid taint flows.

Linvail~\cite{linvail} is a dynamic shadow execution framework based on
source-code-level instrumentation that can be used to implement dynamic taint tracking.
To make sure the analyzed program is not affected by wrapped values,
Linvail permanently unwraps and wraps values around calls and uses
JavaScript proxies to intercept object accesses from non-instrumented code.
However, in order to handle side-effects and keep track of taint labels
in non-instrumented code, it relies on an oracle of hard-coded models.
Providing a complete oracle that preserves the semantics
of the program is known to be very challenging.

Jalangi1~\cite{jalangi1} (not maintained anymore) is another dynamic analysis framework
based on source code-level instrumentation that provides shadow execution to run
different types of analysis, e.g., dynamic taint tracking.
To support primitive values in the shadow execution,
it has hard-coded wrapping and unwrapping operations for language-level
operations, e.g., assignments. For external code that is not instrumented,
its partial solution handles built-in calls that expect primitive
values but receive wrapped values instead. However, it might break the semantics
of the program when wrapped objects reach non-instrumented code. Also, during the offline mode in Jalangi1, external calls
are replaced with concrete values, thereby ignoring side-effects
and potentially deviating from the dynamic execution.
Unlike Jalangi1, Jalangi2~\cite{jalangi2} only provides syntactic traps
 to implement dynamic analyses to avoid the existing problems in Jalangi1.

Compared to the dynamic taint tracking approaches, our taint
inference analysis is based on source code-level instrumentation.
We use Jalangi2 to instrument
JavaScript programs with our analysis code because
it is easy to maintain and can work across different JavaScript engines.
Our analysis is lightweight and can deal with the non-instrumented parts:
built-ins, browser APIs and DOM functions.

Affogato~\cite{affogato} is an instrumentation-based dynamic taint inference analysis
tool for Node.js applications. Similar to \tool, it finds injection vulnerabilities by detecting flows
of data from untrusted  security-sensitive sources to sinks at
runtime using a non-intrusive grey-box taint inference analysis. \tool goes one step further and improves the precision
by introducing a multi-staged approach.

\section{Feedback-driven, guided, and security-aware crawling of modern web applications}
\label{sec:design}

Most of the existing crawlers are designed as standalone
drivers to interact with an application with the goal of getting maximum
coverage of the executed code or discovered hyperlinks.
However, the search space in modern
web applications is simply too large for such a coarse-grained
approach to be useful in practice.
As a result, based on a target analysis,
the security analysts often need to manually identify
which parts of the application should be prioritized and
explored more deeply to find security vulnerabilities.

In this work, we take the first step to automate
guiding a state-aware crawler based on the requirements
of a target security analysis using a feedback-driven approach.
Algorithm~\ref{alg:main-feedback} shows the main
feedback loop of our crawler. This algorithm takes
as input the $URL$ of a web application, a target security analysis, $TA$,
and a set of program locations, $Loc$, that the crawler
should be guided towards.
The loop continues until the crawler reaches a fixpoint and
there are no more new states to visit, and the results
of the target analysis ($Results$) is reported as output.

At each iteration, we run a target security analysis, $TA$,
over the JavaScript and HTML code in the given state, $S$,
and collect results. At the same time, we run
an approximate call graph analysis~\cite{acg} on the
newly discovered JavaScript code, and collect the execution
trace using lightweight instrumentation.
The execution trace helps us to determine the actual (true positive)
function calls, thereby removing false positive call graph edges, and adding
newly discovered ones to the $ACG$, which contains the call graph for
the explored parts of the application.
$PrioritizeEvent(S, ACG)$ determines which
event should be triggered next based on the metrics described
in Sec.~\ref{sec:prioritization} that are computed using $ACG$
and the current state $S$.
In Sec.~\ref{sec:taintinference},
we describe a novel taint inference analysis to detect
DOM-based XSS vulnerabilities as an example security analysis
that fits well in our approach.

\begin{algorithm}
\caption{Feedback-driven and guided security-aware crawler}
\label{alg:main-feedback}
\begin{algorithmic}[1]
  \State inputs: web application $URL$, target analysis $TA$, target program locations $Loc$ 
  \State output: $Results$
  \State $ACG \gets \emptyset$
  \State $browser.goto(URL)$
  \While {$browser.newStateExists()$}
  \State $S \gets browser.getNewState()$
  \State $Results \gets Results~\cup$ \Call{analyze}{$S, TA$} // See Algorithm~\ref{alg:value-fuzzing} 
  \State $cg \gets computeACG(S)$
  \State $trace \gets getExecutionTrace(S)$
  \State $ACG \gets refineACG(ACG, cg, trace)$
  \State $e \gets prioritizeEvent(S, ACG)$
  \State $browser.goto(e)$
  \EndWhile
  \State $report(Results$
\end{algorithmic}
\end{algorithm} 

\subsection{State representation and comparison}
The first key challenge in designing a state-aware crawler is
to come up with a suitable state representation.
Ideally, we would like to store the entire browser
and server state to be able to switch back and forth
between the states gracefully. However, keeping such
a huge amount of information in each state is unrealistic.
To address this challenge, we try
to keep the size of the states minimal by storing only:
(1) URL; and (2) DOM tree, which we have found
to be the most crucial elements.
We also record references to the parent and child states to be able to replay the sequence
of events that have been triggered to reach the current state.

We determine whether a state has been visited before
using the following heuristics: (1) the path segment
of the URL is the same; (2) the size of the DOM tree
has not changed dramatically (less than a threshold);
(3) the hash computed for the DOM tree is exactly the
same, or the difference in DOM tree structure is less than
a threshold.

\subsection{Search strategy}
We perform Depth First Search (DFS)
on the crawler state graph to explore the states. However,
because we do not keep the entire
browser state, it is not possible
to directly backtrack and take the browser
to a previously visited state.
One way to go back to a visited state
is to replay event sequences all the way
from the root to reach a particular state.
This approach requires triggering many
unnecessary events that can significantly affect the efficiency
of the crawler.

To deal with this challenge, we apply heuristics
to take the browser to a previously visited state:
(1) check whether the target state can be 
reached by triggering another event from the 
current state; and (2) compute the
shortest path in the state graph to reach the target state.

\subsection{Call graph refinement}

\begin{listing}[t]
\begin{minted}{html}
<!DOCTYPE html>
<html lang='en'>
<script src='./knockout-min.js'></script>
<script>
function event_handler() {
    fetch('');
}
</script>
<body>
    <button id='knockout_data_bind_1' data-bind='click: button_click'>knockout_data_bind</button>
    <script>
    function KnockoutViewModel() {
        this.button_click = function() {
            event_handler();
        };
        this.button_text = 'knockout_data_bind';
    }
    ko.applyBindings(new KnockoutViewModel());
    </script>
</body>
</html>
\end{minted}
	\caption{An example framework code that is hard for ACG to analyze soundly, and misses a critical call graph edge. Our call graph refinement approach, however, is able to detect and add it to the call graph.}
	\label{lst:cg-example}
\end{listing}

$ComputeACG$ 
in Algorithm~\ref{alg:main-feedback} generates
an approximate call graph using ACG~\cite{acg} for the JavaScript code
executed in the state $S$. Note that the call graph construction
is initially performed statically. As the crawler interacts
with the user interface, we collect the function calls
as part of the execution trace. This execution trace is
next processed to examine whether an edge in the call graph is missing
or is a false positive. The call graph is updated with this new
information.
 An edge from node $a$ to node $b$ in the
call graph is considered as a false positive
if visiting $a$ does not result in visiting $b$. 
And an edge from node $a$ to node $b$ is missing
if the call graph does not include such an edge
but the execution trace does. Listing~
\ref{lst:cg-example} is a code-snippet from the
knockout.js~\cite{knockout} framework. Due to
a complex event delegation mechanism, ACG fails
to find the edge from the \codett{click} event
in the \codett{button} element to \codett{event\_handler}
function. However, once the crawler clicks on this button,
the execution trace records that \codett{event\_handler} gets
triggered. This newly found edge is added
to the call graph, $ACG$.

\subsection{Prioritization}
\label{sec:prioritization}

We prioritize a state that is visited for the first time (not similar to
any of the previously visited states) if it contains
a target program location ($Loc$ in Algorithm~\ref{alg:main-feedback}).
For partially expanded states, we use a prioritization heuristic to choose the next event in the state
that should be triggered by the browser. To guide
the crawler towards target program locations,
we prioritize an event if it has the minimum distance from the handler (registered
to handle it) to a target location in the call graph.

\subsection{Input value generation}
\label{sec:input-gen}

While the state-aware crawler interacts with the JavaScript application,
we analyze (line 7 in Algorithm~\ref{alg:main-feedback})
the JavaScript code returned from the server side\footnote{The JavaScript code is stored in the crawler state.},
generating input values to bypass guards and increase coverage. There are several
ways to provide input values into a client-side JavaScript application:
forms, URLs, cookies, local storage, etc.
In this section, we show how we integrate input value generation for URLs
to our state-aware crawler for simplicity.
However, the same algorithm can be used
for other sources of input values.

The input value generation is performed only on the states
that are candidates to be analyzed by the target analysis ($TA$ in Algorithm~\ref{alg:main-feedback}).
A state is analyzed if its code contains a target program
location.
Algorithm~\ref{alg:value-fuzzing} shows how we generate input values.

At high level,
to bypass the guards on the execution path we collect
runtime values of interest during
the execution, construct path constraints, and
solve them  to generate inputs ($generateNewURLs$ at line 19).
The guards that we aim to bypass are validation routines that must
be satisfied to let the analyzer reach the deeper parts of the program.
Example runtime values of interest are operands in conditional statements (e.g., \codett{if} statement)
and arguments in \codett{string} function calls (e.g., the \codett{string.substring} built-in function)
that are triggered on the execution path.
We use such logged values in constraint generation if they are tainted
(the taint analysis is explained in detail in Sec.~\ref{sec:taintinference}).
These constraints are used to replace tainted characters of input values that are compared in
a conditional statement.

\begin{listing}[ht]
    \inputminted[xleftmargin=1em,linenos]{html}{code-snippets/running-example.tex}
    \caption{An example of HTML/JavaScript code with constraints on input value.}
    \label{lst:constraint-example}
\end{listing}

We explain the input value generation in Algorithm~\ref{alg:value-fuzzing}
through an example. Listing~\ref{lst:constraint-example} shows a simplified
application that is vulnerable to DOM-based XSS attack.
In a DOM-based XSS
attack, the attacker-controllable input (e.g., URL)
is loaded in the victim's browser and the injected payload
flows to a DOM manipulation statement that results in running the malicious payload,
such as stealing cookies, under the victim's session.
In this example, a value obtained from the URL
at line 12 is written to the \codett{document} object at line 14, which
modifies the DOM and allows
to run the attackers malicious payload. However,
the original URL used to load the page is \codett{"http://example.com\#action"},
which does not contain \codett{"show"}. Therefore, line 14 is not executed
when the original input value is used.
Next, we show how we generate an input (URL)
that bypasses the validation at line 13 and allows the execution to reach line 14.

Initially, the execution path ($\pi$) in Algorithm~\ref{alg:value-fuzzing} is empty and the test
input queue, $InputQ$, contains the original URL.
Our input generator continues generating new test inputs
until  $InputQ$ is empty.
In each iteration, an input is removed
from $InputQ$ and passed to the $runTargetAnalysis$
function, which runs the target analysis ($TA$) determined by the analyst.
Before running the target analysis, we take
the browser to state $S$ by obtaining
and triggering the corresponding event sequence ($eventSeq(S)$).

As the target analysis is performed at line 7,
the conditional statements (e.g., \codett{if} statements)
are logged in $\pi$, which are used to generate path constraints
and new test inputs at line 8. Going back to the example in
Listing~\ref{lst:constraint-example}, when analysis executes line 13,
we record the \codett{if} statement
together with the following runtime values in $\pi$: \codett{"show"},
\codett{-1} and \codett{"http://example.com\#action"} (value of \codett{loc} variable).

The $ValueInputGen$ function in Algorithm~\ref{alg:value-fuzzing}
generates new inputs using the values recorded in the execution
path, $\pi$. If a value in a
conditional statement or \codett{string} function call is
identified to be tainted by taint
analysis (See Sec.~\ref{sec:taintinference} for details), the tainted characters
and the value that they are compared against are recorded
in the $Constraints$ map. For instance,
the value of \codett{loc} at line 13 in Listing~\ref{lst:constraint-example}
is inferred to be tainted by taint analysis, and the tainted
characters are \codett{"action"}. Therefore,
\codett{"action"} is added to $Constraints[3].taintedVal$ at line
15 in Algorithm~\ref{alg:value-fuzzing}. We also record \codett{"show"}, which
is the value that the tainted value is compared against.

Finally, the $genConstraint$ function at line 16 in the algorithm generates the
\codett{loc == "show"} constraint and stores it in the $Constraints$ map.
The $generateNewURLs$ function at line 19 replaces
\codett{"action"} with \codett{"show"} in the original
URL\footnote{\codett{http://example.com\#action}}
and generates a new input\footnote{\codett{http://example.com\#show}}.
If the target analysis is DOM-based XSS detection,
once the new URL is loaded and analyzed at line 6 in
Algorithm~\ref{alg:value-fuzzing}, line 14 in
Listing~\ref{lst:constraint-example} is executed and
a DOM-based XSS vulnerability is reported.

\begin{algorithm}
  \caption{Input value generation}
  \label{alg:value-fuzzing}
  \begin{algorithmic}[1]

  \Function{analyze}{$S$, $TA$}
  \State $\pi \gets \emptyset$ // JavaScript execution path
  \State $InputQ \gets URL$ //initial seed input value
  \While{$ not InputQ.isEmpty()$}
  \State $v \gets InputQ.pop()$
    \State $browser.Goto(eventSeq(S)$
    \State $\pi \gets runTargetAnalysis(v, TA)$
    \State $InputQ.add(\Call{ValueInputGen}{\pi})$
  \EndWhile  
  \EndFunction
  
  \Function{ValueInputGen}{$\pi$}
  \State $Constraints \gets \emptyset$
  \For{$n$ in $\pi$}
      \If{$taintAnalysis(n.val)$}
        \State $Constraints[n.loc].taintedVal = taintedVal(n.val)$
        \State $Constraints[n.loc].cons = genConstraint(n)$
      \EndIf    
  \EndFor
  \State Return $generateNewURLs(Constraints, URL)$
  \EndFunction

  \vspace{1em}

\end{algorithmic}
\end{algorithm}

\subsection{Support for other features}
In this section, we elaborate
on the remaining features mentioned in Table~\ref{tbl:comparison}
that are supported by \tool.

\noindent \textbf{Dynamic event handler registration.} We perform dynamic
analysis to collect runtime values, traces and to capture events
that are difficult to detect statically. To perform
the dynamic analysis, we instrument the JavaScript
code to collect traces and hook into the event handlers to capture
the events that otherwise cannot
be found at runtime.\footnote{These are the events that are
registered using \codett{addEventListener}} There are
also many cases where the event registration is done in a library through
a complex mechanism that is not straightforward
(e.g., the event delegation mechanism in \codett{jQuery}~\cite{jquery}).
In such cases, we use models for the common libraries and frameworks
that allow us to extract the required information from their 
 internal data storage.

\noindent \textbf{Static link extraction.} Similar to most of the existing
crawlers, we statically extract links from the HTML pages.
In addition, we integrate static
link extraction to the state prioritization, as follows.
Before the state-aware crawling starts, \tool statically
extracts links from HTML pages starting from $URL$.
This step allows us to retrieve a partial and
coarse-grained structure of the web application. We also build
call graphs statically for the JavaScript code of the extracted
pages and prioritize them for the state-aware crawling
if they contain user-specified target locations ($Loc$ in algorithm~\ref{alg:main-feedback}).

\noindent \textbf{Filling forms.} \tool fills forms
with payloads provided as configuration. Furthermore,
we use the input value generation technique described in Sec.~\ref{sec:input-gen}
to fill form fields.

\section{Target Security Analysis: DOM-based XSS Detection}
\label{sec:taintinference}

In this section, we describe a novel DOM-based XSS detection technique
as an example security analysis that can be integrated
into our guided crawler to detect vulnerabilities in
real-world JavaScript applications. For DOM-based XSS analysis, the crawler needs
to be guided towards DOM manipulation locations, which are
marked as sinks. Once the crawler reaches the states
that contain such sinks, we perform taint analysis
as described below to detect vulnerabilities.

Dynamic taint analysis is a common technique to detect injection vulnerabilities 
in applications written in dynamic languages.  
In practice, however, several technical
challenges must be addressed to implement a dynamic taint analysis that will be able to
analyze real-world applications.
In this section, we first elaborate on the technical challenges
that must be addressed to implement a dynamic taint analysis for JavaScript. Then we present
our dynamic taint inference approach that overcomes or circumvents these challenges.

Dynamic analysis for JavaScript has a short but rich
history~\cite{Andreasen2017} and we can already
extract valuable lessons from existing work.
This type of analysis 
requires instrumentation either at JavaScript engine \cite{lekies, vogt}
or source-code level 
\cite{jalangi1,dexterjs, jstaint} to be carried out.
An engine-level
instrumentation-based analysis would require substantial effort
in order to support multiple engines and to be maintained in the long-term.
Being engine-agnostic is important for security analysis because
it is possible for an attack to work on one engine but not on another one~\cite{dexterjs}.
The dynamic taint analysis should find taint flows
regardless of the engine it is running on.

Previous works show
that source-code level instrumentation-based dynamic taint analysis for JavaScript will face the
following challenges:

\begin{enumerate}
  \item Tracking taint through non-instrumented code. 
  \item Attaching taint labels to primitive values. 
\end{enumerate}

Because modern JavaScript engines are typically implemented in low-level languages (C, C\texttt{++}),
an instrumentation-based dynamic analysis will not be able to instrument all built-in
functions (e.g., array modification, and string operations). Furthermore, for efficiency reasons,
it is often desirable not to instrument an entire application to leave some modules uninstrumented.
Consequently,
the analysis needs to model what happens in non-instrumented code and
update taint labels accordingly. A common approach to deal with non-instrumented code
is to use manually created models~\cite{linvail, dexterjs, jstaint}. Unfortunately, our experience with these tools suggests that their models contain bugs and are incomplete.

Finally, because primitives cannot be extended with additional properties, instrumentation-based 
approaches need to \textit{wrap} primitives in an object that will have a property 
representing the taint label of the primitive value -- aka, boxing. Because wrapping primitives is
an intrusive process that alters the execution of the original program, care must be
taken not to alter its semantics. However, empirical 
evidence from previous work~\cite{linvail, dexterjs, jalangi1} suggests that wrapping primitives while preserving
the semantics of the original program is extremely challenging. In general,
wrapped primitives must be unwrapped before they exit
the instrumented code, and wrapped before they enter
the instrumented code. We suspect that
because of these difficulties primitive wrapping has been dropped
in the new version of Jalangi (Jalangi2)~\cite{jalangi2}.

To highlight the challenges encountered by existing dynamic taint tracking solutions,
consider the code-snippet in Listing~\ref{lst:taint-example}. This example shows a JavaScript
program that contains a vulnerable taint flow
 from the source, \codett{location.hash} at line 1,
 to the security sensitive sink, \codett{document.write} at line 5.
 In this example, the string value \codett{\#payload} is injected in the URL as the fragment identifier (i.e., the part of
the URL following the \codett{\#} sign).
 DexterJS~\cite{dexterjs} fails to report the vulnerable taint flow due to
 an incorrect model used for the uninstrumented built-in
 function, \codett{substring} at line 2.
 On the other hand, Linvail~\cite{linvail} and Chromium Taint Tracking~\cite{lekies} report two taint flows
 at both lines 4 and 5, even though \codett{document.write} at line 4
 is not tainted.

 \begin{listing}[t]
\begin{minted}{javascript}
var tmp = document.location.hash;    // tmp = "#payload"
tmp = tmp.substring(3,7);     // tmp = "yloa"
tmp = tmp + "123";          // tmp = "yloa123"
document.write(tmp.substring(5)) // "23" is written to DOM
document.write(tmp);        // "yloa123" is written to DOM
\end{minted}
	\caption{Example JavaScript program that contains a vulnerable taint flow.}
	\label{lst:taint-example}
\end{listing}

 \subsection{Dynamic Taint Inference}

To address the challenges listed above, we developed a non-intrusive, dynamic taint 
inference analysis based on source code-level instrumentation. Our analysis infers tainted flows by correlating values at sources and sinks, and
observing the behavior of the program instead of attaching and tracking taint labels.



\begin{figure}
\centering
\includegraphics[height=0.5\textheight, width=0.4\textwidth]{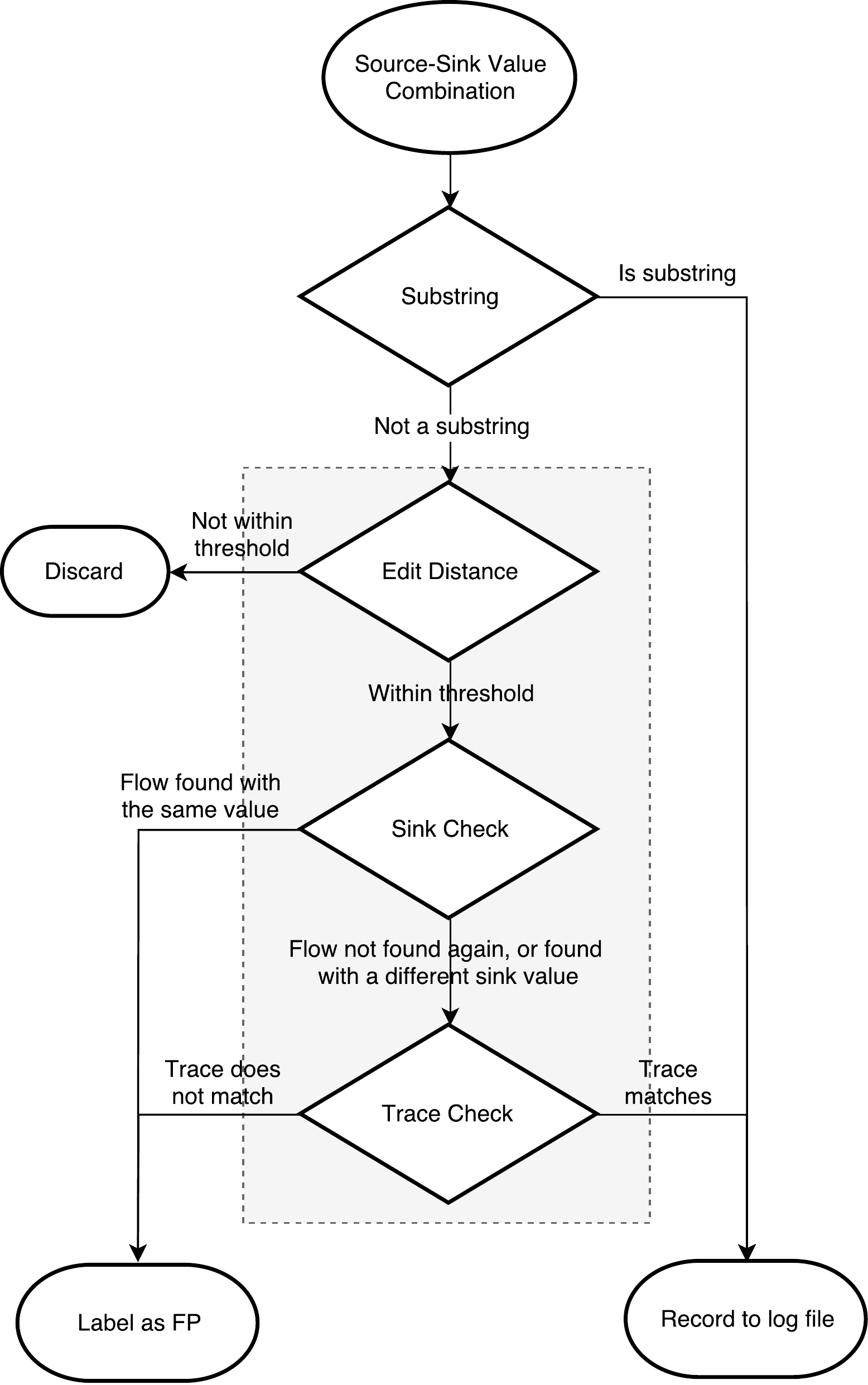}
\caption{Flow diagram of our staged dynamic taint flow inference technique.}
\label{fig:taint_inference_flow}
\end{figure}

Fig.~\ref{fig:taint_inference_flow} shows our staged approach to
discover correlations between values at taint sources and sinks. The stages,
represented as diamonds in Fig.~\ref{fig:taint_inference_flow}, act as increasingly
complex filtering steps that aim to maximise the precision of our approach. 

\subsubsection{Stage 1: Substring}
The first stage looks for an exact substring match of length $\geq \theta$
between string values observed at sources and sinks, i.e., whether either string is a substring of the other. If a match of length
$\geq \theta$ is found, a taint flow is immediately reported. 
Otherwise, the remaining three stages (shown in the grey box) are used to
infer taint flows when the values at sources and sinks approximately match. 
The remaining three stages will be described in the following sections using the 
symbols outlined below.

\begin{itemize}
\item $A = $ a source (identified by location in source-code)
\item $B = $ a sink (identified by location in source-code)
\item $A_v = $ string value at source $A$
\item $B_v = $ string value at sink $B$
\item $F = $ a taint flow detected by our analysis
\end{itemize}

\subsubsection{Stage 2: Edit Distance}
\label{sec:editdistance}
If neither $A_v$ or $B_v$ is a substring of the other (i.e., the substring stage
does not report a match), the edit distance filter
performs approximate matching of $A_v$ and $B_v$. Specifically, this stage
computes the \textit{longest common subsequence} (LCS)~\cite{lcs} between $A_v$ and $B_v$, 
extracts $D_i$ and $D_d$, the number of insertions and deletions required to compute
the LCS, computes a similarity score, and compares it to a threshold $\eta$, as shown 
in Algorithm~\ref{alg:editdistance}.
The source-sink ($A$, $B$) pairs that pass this test in this stage are recorded 
and processed in the next stages to filter out false positives (FPs).

\begin{algorithm}
\caption{Edit Distance Stage}
\label{alg:editdistance}
\begin{algorithmic}[1]
  \Function{EditDistance}{$A_v$, $B_v$, $\eta$}
  \State Let $D_i, D_d = LCS(A_v,B_v)$
  \State Let $L = max(len(A_v), len(B_v))$
  \If{ $\frac{L-(D_i+D_d)}{L} \geq \eta$}
  \State \Return \codett{"Yes"}
  \EndIf
  \State\Return \codett{"No"}
  \EndFunction
\end{algorithmic}
\end{algorithm}

\subsubsection{Stage 3: Sink Check}
According to the flow diagram in Fig.~\ref{fig:taint_inference_flow}, when a 
source-sink pair ($A$, $B$) reaches the sink check stage, we know that 
there is no exact substring match but that the two strings are \textit{similar}.
To weed out cases where the similarity happens by chance (i.e., there is no taint
flow from $A$ to $B$), the sink check stage mutates $A_v$ into $A_v'$ by changing a few characters randomly, running the 
program again with the new source input, and observing $B_v'$. There are three possible
outcomes, as shown in Algorithm~\ref{alg:responsechecking}:

\begin{enumerate}
  \item Sink $B$ is not reached ($B_v'$ is \codett{NULL}). The execution path 
    triggered by $A_v'$ has diverged from the execution path triggered by $A_v$.
    The pair ($A$, $B$) proceeds to the next stage.
  \item $B_v'$ is different from $B_v$, indicating that the value at $A$ has an
    impact on the value at $B$. The pair ($A$, $B$) proceeds to the next stage.
  \item $B_v'$ is identical to $B_v$, indicating that the value at $A$ probably has
    no impact on the value at $B$. The pair ($A$, $B$) does not proceed to the next 
    stage.
\end{enumerate}

\begin{algorithm}[H]
 \caption{Sink Check Stage}
 \label{alg:responsechecking}
 \begin{algorithmic}[1]
   \Function{SinkCheck}{$(A,B)$}
   \State Let $A_v' = mutate(A_v)$
   \State $B_v'$ = runApplication($A_v'$)
   \If{$B_v'$ is NULL}
   \State \Return \codett{"Proceed to next filter"}
   \ElsIf{ $B_v' \neq B_v$}
    \State \Return \codett{"Proceed to next filter"}
   \ElsIf{$B_v' == B_v$}
   \State\Return \codett{"No taint flow from $A$ to $B$"}
   \EndIf
   \EndFunction
 \end{algorithmic}
 \end{algorithm}

\subsubsection{Stage 4: Trace Check}
Trace Check is the final and most expensive stage of our taint flow inference process. 
It aims at detecting real taint flows with high precision. This step involves recording the 
JavaScript execution trace and analyzing the string manipulation operations performed on $A_v$ 
to determine whether $B_v$ is derived from $A_v$ (i.e., there is a taint flow from $A$ to $B$).

Algorithm~\ref{alg:trace} shows our trace check stage. Given a source value $A_v$, a number
of insertions $D_i$, a number of deletions $D_d$, and the execution trace seeded with $A_v$, 
the \textsc{TraceCheck} procedure determines whether the string operations
in the trace can possibly transform $A_v$ into $B_v$. The sub-procedure \textsc{IsOpTainted} 
in Algorithm~\ref{alg:trace} re-uses the Substring and Edit Distance stages, parameterised 
with $\theta$ and $\eta$, to determine whether the base variable or any argument of a string 
operation matches $A_v$. If the base variable or any argument matches $A_v$, \textsc{IsOpTainted} 
returns true.

\begin{algorithm}
  \caption{Trace Check Stage}
  \label{alg:trace}
  \begin{algorithmic}[1]
    \Function{TraceCheck}{$A_v$, $D_i$, $D_d$, $trace$}
    \State Let $D_{ti}$ = 0 and $D_{td}$ = 0
    \For{each $string_{op}$ in $trace$}
      \If{\Call{IsOpTainted}{$string_{op}$, $\theta$, $\eta$, $A_v$}}
        \If{\Call{IsOpInsertion}{$string_{op}$}}
          \State  $D_{ti}$ += 1
        \ElsIf{\Call{IsOpDeletion}{$string_{op}$}}
          \State  $D_{td}$ += 1
        \EndIf
      \EndIf
    \EndFor
    \If{ ($D_{i} >$  0 \&\& $D_{ti}$ == 0) $||$  ($D_{d} >$  0 \&\& $D_{td}$ == 0)}
      \State\Return \codett{"Trace does not match"}
    \Else
      \State\Return \codett{"Trace matches"}
    \EndIf
    \EndFunction   
  \end{algorithmic}
\end{algorithm}

The trace check stage counts the number of tainted string operations that are insertions
($D_{ti}$), and deletions ($D_{td}$) in the trace. Then, it weeds out traces where either no
insertion happens while $D_i > 0$ or no deletion happens while $D_d > 0$.

We now revisit the example in Listing~\ref{lst:taint-example} to show how our taint inference technique
correctly reports an inferred taint flow at line 5 and does not report any flows at line 4.
For the sink at line 4, the observed values are $A_v =$ \codett{"\#payload"} and $B_v =$ \codett{"23"}.
Since $A_v$ and $B_v$ do not pass the check at Substring stage, they are passed to the
Edit Distance stage. As these values also fail to pass the Edit Distance check,
the analysis does not infer any taint flows.

For the sink at line 5 in this example, $A_v =$ \codett{"\#payload"}, $B_v =$ \codett{"yloa123"}, 
$D_i = 3$, $D_d = 4$. Inspecting the trace between line 1 and 4 in this example, the Trace Checking filter is able to 
determine that one string concatenation and two substring operations occurred.
Because both of these operations are performed on the \codett{tmp} base variable with the 
string values \codett{"\#payload"}, \codett{"yloa"} and \codett{"yloa123"}, they pass the \textsc{IsOpTainted}
check at line 4 in Algorithm~\ref{alg:trace} that compares them against the source value \codett{\#payload}. The trace 
check algorithm then computes $D_{ti} = 1$, and $D_{td} = 2$ and concludes that the trace
matches at line 15 in Algorithm~\ref{alg:trace}.

\section{Implementation}
\label{sec:implementation}

We implemented our security-aware guided crawler
in a tool called \tool. Our crawler interacts with the
application running in the
browser using  Pyppeteer~\cite{pyppeteer},  a 
browser automation framework that communicates via ChromeDevTools protocol.
To guide our input generator towards target locations,
we use the pessimistic mode in the
approximate call graph construction~\cite{acg} to statically
build call graphs. We have developed a new lightweight instrumentor
to carry out the dynamic analysis of JavaScript code for dynamic event
handler registration and call graph refinement. The dynamic analysis
finds dynamically registered events, new pages and missing
edges in the call graph at runtime. The lightweight instrumentor
is also used to collect runtime values and generate constraints
for creating new input values.

The source-sink identification, 
Substring and Edit Distance checkers in the taint flow 
inference system are implemented as an analysis written in JavaScript on top of the
Jalangi2~\cite{jalangi2} analysis framework. The Response Check and Trace 
Check components are written in Python, using 
WebSocket to communicate with the JavaScript part of 
the system.
The taint flow reporting, crawler and event generator are 
implemented in Python using bindings for Pyppeteer~\cite{pyppeteer}.

The taint flow inference framework requires a list of sources and sinks to begin with.
The same sinks are used to guide the crawler for DOM-based XSS detection.
These JavaScript methods and property read/write statements are intercepted using Jalangi 
to record and analyze their values.

\section{Evaluation}
\label{sec:evaluation}

The experiments are performed in Google Chrome browser version 69.0.3494.0
on Ubuntu 16.04 running on VirtualBox 6.0,
Intel i7-7700 CPU @ 3.60GHz x 4 (4 cores assigned to VM) with 4096 MB memory.
To evaluate our feedback-driven edge addition technique, we have manually
created models for complex libraries, jQuery~\cite{jquery}, Knockout.js~\cite{knockout},
and React.js~\cite{react.js} that
help find missing edges in the call graph to improve coverage of target locations.
We compare the results of our edge addition technique with these manually created models.
In our experiments we answer
the following research questions:

\begin{itemize}
\item RQ1: How effective is \tool's feedback-driven and guided crawling technique in terms
  of coverage and performance?
\item RQ2: How does \tool compare against other tools for the number of discovered URLs?
\item RQ3: How effective is \tool's DOM-based XSS detection technique compared to the
  state-of-the-art taint analysis techniques in terms of precision and recall?
\end{itemize}

\subsection{Choosing benchmarks}
One of the challenges we have faced for evaluating \tool is choosing
benchmarks. Existing taint analysis works often evaluate on Alexa top~\cite{alexatop}
websites, however,  we are constrained to fuzz and analyze websites
and applications that are published as open-source projects for comparing analysis tools.
Therefore, we have gathered deliberately vulnerable open-source applications, vulnerable
libraries, our in-house
applications, and micro-benchmarks. Our criteria for choosing benchmarks are to:

\begin{itemize}
\item be realistic, diverse, and use modern technologies
\item use complex libraries and show the capability to detect known CVEs
\item include both single-page and multi-page applications
\item show the accuracy of the analysis by evaluating on relevant micro-benchmarks
\item be tested by the related works if possible
\end{itemize}

We evaluate \tool on two target analyses, for which we have collected different benchmarks.
The first target analysis reports the AJAX calls and URLs
found during exploring the client-side web application, which can be used
for REST API testing of the server side. The second
target analysis is DOM-based XSS detection using taint inference, as descibed in Sec.~\ref{sec:taintinference}.
Details of each set of benchmarks are described in the following sections.

\subsection{RQ1: effectiveness of our feedback-driven and guided crawling technique}

\begin{figure*}[htp]
\newcommand{\linecolora}{black}
\newcommand{\linecolorb}{blue}
\newcommand{\linecolorc}{red}
\newcommand{\linecolord}{orange}
\newcommand{\linecolore}{green!40!black}
\newcommand{\linemark}{o}
\begin{minipage}{0.5\textwidth}
\begin{tikzpicture}
\begin{axis}[
  title={WorkBetter (OJET)},
  xlabel={Minutes Crawled},
  ylabel={Number of Unique AJAX Requests},
  xmin=0, xmax=20,
  ymin=0, ymax=20,
  xtick={0,5,10,15,20},
  ytick={0,5,10,15,20},
  legend pos=south east,
  xmajorgrids=true,
  ymajorgrids=true,
  grid style=dashed,
]
\addplot[
  color=\linecolora,
  mark=\linemark,
  ]
  coordinates {
      (0,0)(5,9)(10,13)(15,13)(20,13)(25,13)(30,13)
  };
\addplot[
  color=\linecolorb,
  mark=\linemark,
  ]
  coordinates {
      (0,0)(5,15)(10,15)(15,18)(20,18)(25,19)(30,20)
  };
\addplot[
  color=\linecolorc,
  mark=\linemark,
  ]
  coordinates {
      (0,0)(5,15)(10,15)(15,15)(20,16)(25,16)(30,16)
  };
\addplot[
  color=\linecolord,
  mark=\linemark,
  ]
  coordinates {
      (0,0)(5,15)(10,18)(15,19)(20,20)(25,20)(30,20)
  };
\legend{Random,ACG + Models,ACG + EA,ACG + Both}
\end{axis}
\end{tikzpicture}
\end{minipage}
\begin{minipage}{0.5\textwidth}
\begin{tikzpicture}
\begin{axis}[
  title={Archivist},
  xlabel={Minutes Crawled},
  ylabel={Number of Unique AJAX Requests},
  xmin=0, xmax=120,
  ymin=0, ymax=100,
  xtick={0,20,40,60,80,100,120},
  ytick={0,20,40,60,80,100},
  legend pos=south east,
  xmajorgrids=true,
  ymajorgrids=true,
  grid style=dashed,
]
\addplot[
  color=\linecolora,
  mark=\linemark,
  ]
  coordinates {
      (0,0)(20,22)(40,26)(60,26)(80,26)(100,26)(120,26)
  };
\addplot[
  color=\linecolorb,
  mark=\linemark,
  ]
  coordinates {
      (0,0)(20,43)(40,58)(60,70)(80,74)(100,78)(120,83)
  };
\addplot[
  color=\linecolorc,
  mark=\linemark,
  ]
  coordinates {
      (0,0)(20,42)(40,53)(60,72)(80,80)(100,87)(120,95)
  };
\addplot[
  color=\linecolord,
  mark=\linemark,
  ]
  coordinates {
      (0,0)(20,37)(40,51)(60,58)(80,64)(100,76)(120,81)
  };
\legend{Random,ACG + Models,ACG + EA,ACG + Both}
\end{axis}
\end{tikzpicture}
\end{minipage}
\begin{minipage}{0.5\textwidth}
\begin{tikzpicture}
\begin{axis}[
  title={WebScanTest},
  xlabel={Minutes Crawled},
  ylabel={Number of Unique AJAX Requests},
  xmin=0, xmax=20,
  ymin=0, ymax=80,
  xtick={0,5,10,15,20},
  ytick={0,20,40,60,80},
  legend pos=north west,
  xmajorgrids=true,
  ymajorgrids=true,
  grid style=dashed,
]
\addplot[
  color=\linecolora,
  mark=\linemark,
  ]
  coordinates {
      (0,0)(5,2)(10,23)(15,26)(20,26)
  };
\addplot[
  color=\linecolorb,
  mark=\linemark,
  ]
  coordinates {
      (0,0)(5,2)(10,25)(15,27)(20,28)
  };
\addplot[
  color=\linecolorc,
  mark=\linemark,
  ]
  coordinates {
      (0,0)(5,2)(10,14)(15,25)(20,44)
  };
\addplot[
  color=\linecolord,
  mark=\linemark,
  ]
  coordinates {
      (0,0)(5,2)(10,27)(15,64)(20,67)
  };
\legend{Random,ACG + Models,ACG + EA,ACG + Both}
\end{axis}
\end{tikzpicture}
\end{minipage}
\begin{minipage}{0.5\textwidth}
\begin{tikzpicture}
\begin{axis}[
  title={Juice Shop},
  xlabel={Minutes Crawled},
  ylabel={Number of Unique AJAX Requests},
  xmin=0, xmax=60,
  ymin=0, ymax=30,
  xtick={0,10,20,30,40,50,60},
  ytick={0,5,10,15,20,25,30},
  legend pos=south east,
  xmajorgrids=true,
  ymajorgrids=true,
  grid style=dashed,
]
\addplot[
  color=\linecolora,
  mark=\linemark,
  ]
  coordinates {
      (0,0)(10,18)(20,18)(30,18)(40,18)(50,18)(60,18)
  };
\addplot[
  color=\linecolorb,
  mark=\linemark,
  ]
  coordinates {
      (0,0)(10,12)(20,14)(30,15)(40,16)(50,17)(60,17)
  };
\addplot[
  color=\linecolorc,
  mark=\linemark,
  ]
  coordinates {
      (0,0)(10,12)(20,14)(30,15)(40,16)(50,16)(60,16)
  };
\addplot[
  color=\linecolord,
  mark=\linemark,
  ]
  coordinates {
      (0,0)(10,14)(20,16)(30,17)(40,18)(50,18)(60,18)
  };
\addplot[
  color=\linecolore,
  mark=\linemark,
  ]
  coordinates {
      (0,0)(10,13)(20,14)(30,15)(40,15)(50,20)(60,27)
  };
\legend{Random,ACG + Models,ACG + EA,ACG + Both,ACG + Random}
\end{axis}
\end{tikzpicture}
\end{minipage}
\caption{Evaluating guided crawling against random crawling.}
\label{fig:charts}
\end{figure*}
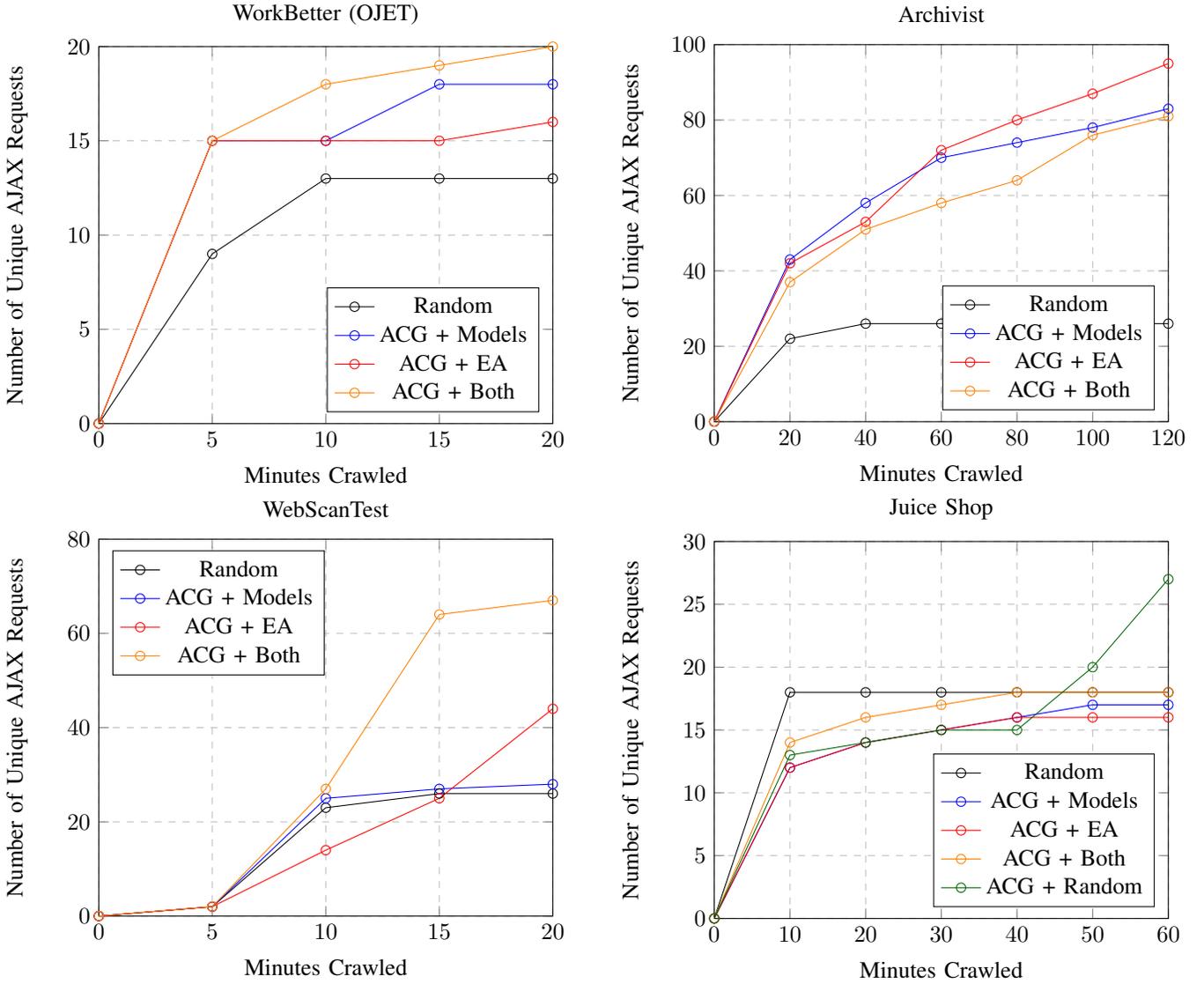

First we evaluate the effectiveness of our
feedback-driven and guided crawling
technique by measuring the performance and coverage
of AJAX calls on two internal Oracle applications,
WebScanTest~\cite{webscantest}, which is the live instance of
a program used to evaluate crawlers, and Juice Shop v8.3.0~\cite{juice-shop}, a modern
and sophisticated deliberately vulnerable web application.
We decided to use these
applications for our experiments because they make use of various
modern technologies and libraries,
such as jQuery~\cite{jquery}, Knockout.js~\cite{knockout},
React.js~\cite{react.js}, and AngularJS~\cite{angularjs}.
To compare the overall coverage
of our approach with other crawlers, we ran
well-known crawlers on open-source
applications that are also used as benchmarks in~\cite{jaak}.

Fig.~\ref{fig:charts} shows how our guided crawling strategy
compares to a non-guided random crawling strategy.
Note that the random strategy still benefits from
our dynamic analysis techniques and only replaces
the state and event prioritization functions in Algorithm~\ref{alg:main-feedback} with
random selection.

In this experiment, we count the number of
distinct AJAX calls made by each strategy
over time. The timeout for this experiment is 120 minutes,
but we don't show the results once all the strategies start
to plateau or they reach a fixpoint, i.e., finish running,
before the timeout.
We evaluate three versions of our guided crawling technique:
(1) ACG + Models, which uses manually crafted models
for libraries, such as jQuery when the approximate call graph
fails to analyze them effectively;
(2) ACG + EA, which is our feedback-driven edge addition
technique to refine the statically generated
call graph during runtime execution and add newly
found edges; and (3) ACG + BOTH,
which uses both Models and EA. By comparing these three versions,
we evaluate the effectiveness of our novel ACG Edge Addition technique,
which is fully automatic
and can be used when manually crafted models for libraries are not available.

\noindent \textbf{Archivist:} 
Archivist is a highly interactive application
for document management that uses the AngularJS~\cite{angularjs} framework
and jQuery~\cite{jquery} library.
The random strategy performs significantly worse than our ACG guided strategies
consistently. The ACG + EA
is not as effective as the manually crafted models for this
application but the gap between them is small. Therefore, replacing the
manual models with the edge addition technique can be promising.

\noindent \textbf{WorkBetter:}
WorkBetter is a tutorial application used to demonstrate
the OJET~\cite{ojet} framework. Apart from OJET, it also
uses the Knockout.js~\cite{knockout} framework and jQuery library.
All of the ACG guided strategies outperform the random strategy on this application.
Among the ACG strategies, ACG + BOTH outperforms the others and the EA is more effective
than the manually crafted models. The results on this application show that
not only EA can replace the manual models but also achieves
better coverage within the same amount of time.

\noindent \textbf{WebScanTest:}
The ACG Edge Addition strategy for this benchmark outperforms
the two other strategies for the AJAX call detection experiment.
The ACG Model finds almost the same number of AJAX calls as the random strategy.
ACG + Both outperforms the other strategies
and  EA is more effective
than the manually crafted models towards the end.

\noindent \textbf{Juice Shop v8.3.0:}
Unlike the other applications in this
experiment, the ACG strategies do not outperform
Random, ACG + Both
catching up with Random only after 40 minutes of crawling.
Juice Shop is a highly AJAX-driven application
and many of the events result in triggering
an AJAX call. Therefore, even a Random strategy can be effective due
to the nature of the application. In fact, our closer investigation
revealed that ACG strategies overprioritize certain events in this case,
leaving few chances for other events to be triggered. To
further understand this behaviour, we experimented with a
hybrid strategy (ACG + Random), where one in five events is chosen
randomly and the rest of the events are prioritized based
on the call graph distance metric. The result shows that
this hybrid strategy significantly outperforms the pure
ACG strategies after 45 minutes. In future,
we plan to experiment with such hybrid strategies for
applications that are similar in nature to Juice Shop.

In summary, we show that our guided crawling strategy
finds AJAX calls
more quickly than the random strategy in all three applications,
which make use of modern technologies and are non-trivial to crawl.
Moreover, our  feedback-driven edge addition (ACG EA)
technique is shown to be effective:
the number of AJAX calls found by this technique are in
the same ballpark as the carefully constructed manual models.

In addition to the guided crawling strategy, we
integrate dynamic analysis of JavaScript,
link extraction and state-aware crawling in a novel way.
We compare this novel design against
state-of-the-art crawlers on open-source
applications that are also used as benchmarks in~\cite{jaak}.
Table~\ref{table:crawlers} shows the total number of URLs and AJAX calls
recorded by each tool.
The results show that our crawler records more URLs and
AJAX calls than other tools except for Wivet benchmark (Arachni
finds six more URLs).

\begin{table}[hbtp]
\caption{Comparing our crawler against state-of-the-art crawlers for number of
  URLs and AJAX calls.}
    \centering
{\scriptsize
    \begin{tabular}{|c|c|c|c|c|}
      \hline
      Crawlers & \multicolumn{4}{c|}{Tested Web Applications}\\
      \cline{2-5}
      & Public firing range & DVWA & Wivet & WebGoat\\
      \hline
      Our prototype & 268 & 76 & 77 & 85\\
      \hline
      Arachni~\cite{arachni} & 257 & 27 & 83 & 3\\
      \hline
      Crawljax~\cite{crawljax} & 83 & 20 & 15 & 2\\
      \hline
      Htcap~\cite{htcap} & 160 & 63 & 68 & 65\\
      \hline
      j\"{A}k~\cite{jaak} & 257 & 34 & 76 & 83\\
      \hline

\end{tabular}
}
\label{table:crawlers}
\end{table}

\subsection{Results for DOM-based XSS detection}

For the second target analysis, DOM-based XSS detection, first we evaluate
the effectiveness of our staged taint inference techniques
on micro benchmarks and libraries with known DOM-based XSS vulnerabilities.
We use two open-source micro benchmarks
designed to evaluate DOM-based XSS detection tools: Firing
Range~\cite{firingrange}\footnote{For the  Firing Range
  benchmark, we evaluate only against the DOM-related
  test cases: address tests~\cite{fr1},
  urldom tests~\cite{fr2} and dom tests (toxicdom)~\cite{fr3}.}
from Google, and IBM benchmarks~\cite{ibmtest}.
We also compare our taint inference technique
against dynamic taint tracking in DexterJS~\cite{dexterjs},
and CTT (Chromium Taint Tracking)~\cite{lekies},
the state-of-the-art DOM-based XSS detection tools.
Next, we evaluate
the effectiveness of our guided crawling strategy and
input generation technique for
DOM-based XSS detection analysis.

\subsubsection{Taint flow inference on microbenchmarks}
The inputs to a DOM-based XSS vulnerability are often provided in the URLs, which
can easily be controlled by attackers.
While some of the test cases in the Firing Range and IBM 
benchmarks contain a valid flow from a source to a sink, the value at the 
source cannot be directly tainted through URLs, e.g., \codett{sessionStorage}.
Table \ref{tab:microbenchmarks_controllable} considers only 
test cases that can be triggered through a user-controlled URL, effectively 
also omitting the test cases for which DexterJS and CTT report false positives. The 
criteria used to label test cases as being \textit{controllable} are as follows:
\begin{itemize}
\item Contains at least one valid  URL-controllable input (source) 
that enters the JavaScript program.
\item Contains at least one valid sink.
\end{itemize}

\begin{table}[hbtp]
  \caption{Vulnerability detection on microbenchmarks -- URL-controllable test cases}
  \centering
  {\scriptsize
  \begin{tabular}{|c|c|c|c|c|}
    \hline
    Benchmark & \multicolumn{3}{c|}{Firing Range} & IBM\\
              \cline{2 - 4}
              & address tests & urldom tests & dom tests (toxicdom) & \\ \hline
 \# Test Cases & 28 & 26 & 3 & 66\\ \hline
 \# \tool & 23 & 22 & 3 & 60\\ \hline
 \# DexterJS & 21 & 6 & 3 & 53\\ \hline
 \# CTT & 18 & 10 & 3 & 54\\ \hline
\end{tabular}
}
\label{tab:microbenchmarks_controllable}
\end{table}

Table \ref{tab:microbenchmarks_controllable} shows that our framework finds
more taint flows than DexterJS and CTT.
This is particularly apparent in urldom test cases,
where our framework is able to detect a significantly higher portion of the URL-controllable
test cases. In particular, we can
report the taint flow in \codett{Incorrect\_Sanitizer/apollo\_test\_01.html} test case in the IBM benchmark, which
is triggered if the URL has \codettf{topic=} query parameter. \tool successfully finds this query parameter and reports the taint flow.
This test case shows the effectiveness of our value input generator and constraint heuristics explained in Sec.\ref{sec:input-gen}.

We investigated the results from the other tools further to understand why they fail to report many valid
taint flows. We noticed that DexterJS
does not handle all built-in functions, sanitizations
and browser APIs. On the other hand, CTT does not handle property reads and writes,
which result in false negatives.
We also noticed problems in the event-generation component
of DexterJS that leads to poor coverage and missing valid taint flows. Because
CTT does not have support for event generation,
it misses all the flows that require user interaction. Also,
the sources and sinks in some of the test cases are not supported in these tools.
However, our framework can handle most of these test cases and successfully
reports valid taint flows.

\begin{table}[hbtp]
  \caption{False positives detected on non URL-controllable test cases}
  \centering
  {\scriptsize
  \begin{tabular}{|c|c|c|c|c|}
    \hline
    Benchmark & \multicolumn{3}{c|}{Firing Range} & IBM\\
              \cline{2 - 4}
              & address tests & urldom tests & dom tests (toxicdom) & \\ \hline
 \# Test Cases & 1 & 0 & 33 & 70\\ \hline
 \# \tool & 0 & 0 & 0 & 0\\ \hline
 \# DexterJS & 0 & 0 & 26 & 3\\ \hline
 \# CTT & 0 & 0 & 12 & 6\\ \hline
\end{tabular}
}
\label{tab:microbenchmarks_fp}
\end{table}

Table \ref{tab:microbenchmarks_fp} shows how our framework outperforms the dynamic taint tracking used in the other tools for false positives.
These results show that the false positive reduction strategies discussed in Sec.~\ref{sec:taintinference} 
are very effective in lowering false positive rate and improving precision. However, whether all the 33 dom tests
are actually false positives is debatable. The taint source in these test cases are not URL controllable and are resources
such as cookies and \codett{localStorage} that are set by the application if they are empty. Therefore, if they are not set,
the application sets them to benign values.  Listing~\ref{lst:fp-example} shows
an example where the \codett{badValue} item in the \codett{localStorage} is set to a constant value that is benign if it is not set before, hence \tool does not report any taint flows.\footnote{We do inject any payloads to any resources, such as cookies and \codettf{localStorage} in any of the experiments in this paper.} CCT still considers the benign values set to these resources as valid
taint sources (instead of killing the taint). Because the taint analysis
in~\cite{lekies} is followed up by an exploit generation step that confirms the taint flows, their DOM-based XSS tool might not report them as exploitable
vulnerabilities.\footnote{We did not have access to the exploit generation component to confirm.}

\begin{listing}[ht]
    \inputminted[xleftmargin=1em,linenos]{html}{code-snippets/fp-example.tex}
    \caption{An example of HTML/JavaScript code where the taint source is set by a benign constant value if it is empty.}
    \label{lst:fp-example}
\end{listing}

\subsubsection{Taint flow inference on popular JavaScript libraries and open-source applications }
Table~\ref{tab:libraries} reports the effectiveness of our taint inference mechanism
on some of the JavaScript libraries that have known 
vulnerabilities, as reported in RetireJS~\cite{retirejs}. RetireJS 
documents vulnerabilities in ``retired'' versions of 
JavaScript libraries. We have created test harnesses for these libraries that trigger the
DOM-based XSS vulnerable paths.  The payloads tested in these experiments are injected as payloads for all the tools.
 \tool is able to report taint flows
for all of these vulnerable libraries while DexterJS misses all
of them and CTT misses two. Dojo is an
interesting library in our benchmarks because its vulnerability can be found if the
analysis can bypass the input validation.
The code-snippet in Listing~\ref{lst:dojo} shows
a simplified version of the vulnerability in this library. \tool successfully
reports the taint flow from line 6 to 21 by appending \codett{theme} as the query parameter
in the URL and bypassing the input validations at lines 5 and 20 using the constraint heuristics explained in Sec.\ref{sec:input-gen}.

\begin{table}[hbtp]
  \caption{Vulnerability detection on JavaScript libraries with known 
    vulnerabilities. Some of the payloads are injected into the fragment identifier part of the URLs: jQuery 1.11.1 does not require any specific payload to reach the vulnerable sink; \codettf{<img src=/ onerror=alert(1)>} is used for jQuery 1.6.1 and jQuery-migrate 1.1.1 and  \codettf{x" onerror=alert(1) nothing} for handlebars 1.0.0.beta.2 and mustache 0.3.0. The payload for dojo is injected as a URL query parameter and needs a specific key: \codettf{?theme=payload}, which is automatically generated by \tool.}
  \centering
  {\scriptsize
  \begin{tabular}{|c|c|c|c|c|}
    \hline
    Benchmark & \tool &  DexterJS &  CTT \\
    \hline
    jQuery 1.11.1 \cite{cve20159251} & \checkmark & ${\times}$ & ${\times}$  \\
    \hline
    jQuery 1.6.1  \cite{cve20114969} & \checkmark & ${\times}$ & \checkmark\\
    \hline
    jQuery-migrate 1.1.1 \cite{jquery-migrate} & \checkmark & ${\times}$ & \checkmark \\
    \hline
    handlebars 1.0.0.beta.2 \cite{handlebars} & \checkmark & $\times$ & \checkmark \\
    \hline
    mustache 0.3.0 \cite{mustache} & \checkmark & $\times$ & \checkmark \\
    \hline
    dojo 1.4.1 \cite{dojo} & \checkmark & $\times$ & $\times$ \\
    \hline
  \end{tabular}
  }
\label{tab:libraries}
\end{table}

Table~\ref{tab:vulnerable-apps} shows that \tool can successfully detect
DOM-based XSS vulnerabilities in non-trivial modern web-applications, while the other tools
miss reporting them. To detect the vulnerabilities in these applications
a state-aware crawler is needed that triggers the vulnerable path. In this experiment, \tool is the only tool
that has an effective state-aware crawler, successfully exploring the application
and triggering the vulnerable path.

\begin{table}[hbtp]
  \caption{Open-source deliberately vulnerable applications.}
  \centering
  {\scriptsize
  \begin{tabular}{|c|c|c|c|c|}
    \hline
    Benchmark & \tool &  DexterJS &  CTT \\
    \hline
    Juice-shop 8.3.0 \cite{juice-shop} & \checkmark & ${\times}$ & ${\times}$  \\
    \hline
    Damn Vulnerable Web App (DVWA)  \cite{dvwa} & \checkmark & ${\times}$ & ${\times}$\\
    \hline
  \end{tabular}
  }
\label{tab:vulnerable-apps}
\end{table}

\subsubsection{Effectiveness of guided crawler and data input generation for DOM-based XSS detection}

We evaluate the effectiveness of guided crawling strategy for
DOM-based XSS detection analysis by comparing
it with the random strategy.
The target locations in this experiment are DOM manipulation
operations in the program. 
Table~\ref{table:guided-dom} shows
that for both of our microbenchmarks,
the guided crawling helps find the DOM-based XSS vulnerabilities
faster than a crawler that uses a random strategy.

\begin{listing}[ht]
    \inputminted[xleftmargin=1em,linenos]{html}{code-snippets/dojo.tex}
    \caption{A simplified snippet from the Dojo version 1.4 library that is vulnerable to DOM-based XSS, where attacker-controllable input enters via \codettf{theme} URL query parameter through the \codettf{window.location.href} property and is written to DOM at line 21.}
    \label{lst:dojo}
\end{listing}

\begin{table}[hbtp]
  \caption{Comparing our guided crawling strategy (ACG) with random strategy (Random) for DOM-based XSS detection.}
    \centering
{\scriptsize
    \begin{tabular}{|c|c|c|c|c|c|c|}
      \hline
      Benchmark & Analysis Step & ACG & Random\\
      \hline
      & Link Extraction & 19m, 51s & 19m, 39s \\
      \cline{2-4}
      Firing Range & State crawling & 3m, 35s & 10m, 12s \\
      \cline{2-4}
      & Taint inference & 10m, 39s & 40m, 18s\\
      \cline{2-4}
      & Total & 34m, 31s & 70m, 27s\\
      \hline
      & Link Extraction & 16m, 41s & 16m, 56s \\
      \cline{2-4}
      & State crawling & 7m, 6s & 6m, 17s \\
      \cline{2-4}
      IBM & Taint inference & 14m, 42s & 52m, 42s\\
      \cline{2-4}
      & Total & 38m, 58s & 76m, 55s\\
      \hline
\end{tabular}
}
\label{table:guided-dom}
\end{table}

\subsection{Threats to validity}
While we aimed to select representative applications and
benchmarks that use modern technologies,
the choice of benchmarks might
have affected the validity of the experiments presented in this paper.
In the experiments we showed that the DOM-based XSS detection in \tool has a high accuracy
for the analyzed applications and libraries. However, depending on the complexity
of the taint manipulation operations in the given program, the accuracy can vary.

\section{Conclusion}
\label{sec:conclusion}

In this paper, we proposed \tool, a dynamic analysis tool
that detects vulnerabilities in
modern and complex client-side JavaScript applications, which are often built upon libraries and frameworks.
We studied the state-of-the-art tools and presented the most crucial
features a security-aware client-side analysis
should be supporting. We proposed the first security-guided client-side analysis
that closes the gap between state-aware crawling and client-side
security analysis by taking a feedback-driven approach
that combines static and dynamic analysis to analyze complex frameworks
automatically, and increase the coverage of security-sensitive
parts of the program efficiently. We evaluated \tool on
various applications and benchmarks with different levels
of complexity, and showed it outperforms the existing crawlers.
Finally, we proposed a new lightweight non-intrusive client-side taint analysis
that reports non-trivial
taint flows on modern JavaScript applications, and has higher accuracy compared to the existing dynamic
client-side taint analysis tools.

\bibliographystyle{plain}
\bibliography{references}

\begin{thebibliography}{10}

\bibitem{alexatop}
{Alexa top}.
\newblock \url{https://www.alexa.com/topsites}.

\bibitem{angularjs}
{AngularJS}.
\newblock \url{https://angularjs.org/}.

\bibitem{arachni}
{Arachni Framework v1.4}.
\newblock \url{http://www.arachni-scanner.com/blog/tag/crawl/}.

\bibitem{cve20114969}
{CVE-2011-4969}.
\newblock \url{http://web.nvd.nist.gov/view/vuln/detail?vulnId=CVE-2011-4969}.

\bibitem{cve20159251}
{CVE-2015-9251}.
\newblock \url{https://nvd.nist.gov/vuln/detail/CVE-2015-9251}.

\bibitem{dvwa}
{Damn Vulnerable Web Application (DVWA)}.
\newblock \url{http://dvwa.co.uk/}.

\bibitem{dojo}
{DOM-Based XSS in Dojo Toolkit SDK}.
\newblock \url{https://bugs.dojotoolkit.org/ticket/10773}.

\bibitem{fr1}
{Firing Range: Address DOM-based XSS}.
\newblock \url{https://public-firing-range.appspot.com/address/index.html}.

\bibitem{fr3}
{Firing Range: DOM tests (toxicdom)}.
\newblock \url{https://public-firing-range.appspot.com/dom/index.html}.

\bibitem{firingrange}
{Firing Range Test Bed}.
\newblock \url{https://public-firing-range.appspot.com/}.

\bibitem{fr2}
{Firing Range: URL-based DOM-based XSS}.
\newblock \url{https://public-firing-range.appspot.com/urldom/index.html}.

\bibitem{handlebars}
{Handlebars.js GitHub Issue \#68}.
\newblock \url{https://github.com/wycats/handlebars.js/pull/68}.

\bibitem{htcap}
{Htcap v1.0.1}.
\newblock \url{https://github.com/segment-srl/htcap/releases}.

\bibitem{jalangi2}
{Jalangi2}.
\newblock \url{https://github.com/Samsung/jalangi2}.

\bibitem{jquery}
{jQuery}.
\newblock \url{https://jquery.com/}.

\bibitem{jquery-migrate}
{jQuery Bug Tracker Issue \#11290}.
\newblock \url{https://bugs.jquery.com/ticket/11291}.

\bibitem{jstaint}
{jsTaint}.
\newblock \url{https://github.com/idkwim/jsTaint}.

\bibitem{juice-shop}
{Juice-shop 8.3.0}.
\newblock \url{https://github.com/bkimminich/juice-shop}.

\bibitem{knockout}
{Knockout.js}.
\newblock \url{https://knockoutjs.com/}.

\bibitem{ibmtest}
{LaBaSec: Language-based Security}.
\newblock
  \url{http://m.ibm.com/http/researcher.ibm.com/researcher/view_group_subpage.php?id=1598}.

\bibitem{mustache}
{Mustache.js GitHub Issue \#112}.
\newblock \url{https://github.com/janl/mustache.js/issues/112}.

\bibitem{ojet}
{OJET}.
\newblock \url{https://www.oracle.com/webfolder/technetwork/jet/index.html}.

\bibitem{owasp_domxss}
{OWASP DOM-Based XSS}.
\newblock \url{https://www.owasp.org/index.php/DOM_Based_XSS}.

\bibitem{zap}
{OWASP ZAP}.
\newblock \url{https://www.owasp.org/index.php/OWASP_Zed_Attack_Proxy_Project}.

\bibitem{pyppeteer}
{Pyppeteer}.
\newblock \url{https://github.com/miyakogi/pyppeteer}.

\bibitem{react.js}
{React.js}.
\newblock \url{https://reactjs.org/}.

\bibitem{retirejs}
{Retire.js}.
\newblock \url{http://retirejs.github.io/retire.js/}.

\bibitem{w3af}
{w3af v1.6.49}.
\newblock \url{http://w3af.org/plugins/crawl}.

\bibitem{webscantest}
{WebScanTest}.
\newblock \url{https://www.webscantest.com/}.

\bibitem{Andreasen2017}
Esben Andreasen, Gong Liang, Michael Pradel, Tu~Darmstadt, Marija Selakovic,
  and Koushik Sen.
\newblock {A Survey of Dynamic Analysis and Test Generation for JavaScript}.
\newblock {\em ACM Computing Surveys}, 2017.

\bibitem{artemis}
Shay Artzi, Julian Dolby, Simon~Holm Jensen, Anders M\o{}ller, and Frank Tip.
\newblock {A Framework for Automated Testing of {J}ava{S}cript Web
  Applications}.
\newblock In {\em ICSE}, 2011.

\bibitem{efficient1}
Kamara Benjamin, Gregor Von~Bochmann, Mustafa~Emre Dincturk, Guy-Vincent
  Jourdan, and Iosif~Viorel Onut.
\newblock {A Strategy for Efficient Crawling of Rich Internet Applications}.
\newblock In {\em ICWE}, 2011.

\bibitem{aflgo}
Marcel B\"{o}hme, Van-Thuan Pham, Manh-Dung Nguyen, and Abhik Roychoudhury.
\newblock {Directed Greybox Fuzzing}.
\newblock In {\em CCS}, 2017.

\bibitem{linvail}
Laurent Christophe, Elisa~Gonzalez Boix, Wolfgang De~Meuter, and Coen
  De~Roover.
\newblock {Linvail: A general-purpose platform for shadow execution of
  JavaScript}.
\newblock In {\em SANER}, 2016.

\bibitem{acg}
Asger Feldthaus, Max Sch\"{a}fer, Manu Sridharan, Julian Dolby, and Frank Tip.
\newblock {Efficient Construction of Approximate Call Graphs for JavaScript IDE
  Services}.
\newblock In {\em ICSE}, 2013.

\bibitem{affogato}
Fran\c{c}ois Gauthier, Behnaz Hassanshahi, and Alexander Jordan.
\newblock {AFFOGATO: Runtime Detection of Injection Attacks for Node.Js}.
\newblock In {\em SOAP}, 2018.

\bibitem{WALA}
Salvatore Guarnieri, Marco Pistoia, Omer Tripp, Julian Dolby, Stephen Teilhet,
  and Ryan Berg.
\newblock {Saving the World Wide Web from Vulnerable JavaScript}.
\newblock In {\em ISSTA}, 2011.

\bibitem{autogram}
Matthias H\"{o}schele and Andreas Zeller.
\newblock {Mining Input Grammars with AUTOGRAM}.
\newblock In {\em ICSE-C}, 2017.

\bibitem{artemisSID}
Casper~Svenning Jensen, Anders M{\o}ller, and Zhendong Su.
\newblock {Server interface descriptions for automated testing of JavaScript
  web applications}.
\newblock In {\em ESEC/FSE}, 2013.

\bibitem{TAJS}
Simon~Holm Jensen, Anders M{\o}ller, and Peter Thiemann.
\newblock {Type Analysis for JavaScript}.
\newblock In {\em SAS}, 2009.

\bibitem{SAFE}
Hongki Lee, Sooncheol Won, Joonho Jin, Junhee Cho, and Sukyoung Ryu.
\newblock {SAFE: Formal specification and implementation of a scalable analysis
  framework for ECMAScript}.
\newblock In {\em FOOL}, 2012.

\bibitem{lekies}
Sebastian Lekies, Ben Stock, and Martin Johns.
\newblock {25 million flows later: large-scale detection of DOM-based XSS}.
\newblock In {\em CCS}, 2013.

\bibitem{domsday}
William Melicher, Anupam Das, Mahmood Sharif, Lujo Bauer, and Limin Jia.
\newblock {Riding out DOMsday: Towards Detecting and Preventing DOM Cross-Site
  Scripting}.
\newblock In {\em NDSS}, 2018.

\bibitem{crawljax}
Ali Mesbah, Arie van Deursen, and Stefan Lenselink.
\newblock {Crawling Ajax-Based Web Applications Through Dynamic Analysis of
  User Interface State Changes}.
\newblock {\em ACM Trans. Web}, 2012.

\bibitem{atusa}
Ali Mesbah, Arie van Deursen, and Danny Roest.
\newblock {Invariant-Based Automatic Testing of Modern Web Applications}.
\newblock {\em IEEE Trans. Softw. Eng.}, 2012.

\bibitem{feedex}
Amin Milani~Fard and Ali Mesbah.
\newblock {Feedback-directed exploration of web applications to derive test
  models}.
\newblock In {\em ISSRE}, 2013.

\bibitem{lcs}
S~Needleman and C~Wunsch.
\newblock {A general method applicable to the search for similarities in the
  amino acid sequences of two proteins}.
\newblock In {\em J. Mol. Biol. 48}, 1970.

\bibitem{survey2010}
Christopher Olston and Marc Najork.
\newblock {Web Crawling}.
\newblock {\em Found. Trends Inf. Retr.}, 2010.

\bibitem{dexterjs}
Inian Parameshwaran, Enrico Budianto, Shweta Shinde, Hung Dang, Atul Sadhu, and
  Prateek Saxena.
\newblock {Auto-patching DOM-based XSS at scale}.
\newblock In {\em FSE}, 2015.

\bibitem{jaak}
Giancarlo Pellegrino, Constantin Tsch{\"{u}}rtz, Eric Bodden, and Christian
  Rossow.
\newblock {j{\"{A}}k: Using Dynamic Analysis to Crawl and Test Modern Web
  Applications}.
\newblock In {\em RAID}, 2015.

\bibitem{vuzzer}
Sanjay Rawat, Vivek Jain, Ashish Kumar, Lucian Cojocar, Cristiano Giuffrida,
  and Herbert Bos.
\newblock {VUzzer: Application-aware Evolutionary Fuzzing}.
\newblock In {\em NDSS}, 2017.

\bibitem{kudzu}
Prateek Saxena, Devdatta Akhawe, Steve Hanna, Feng Mao, Stephen McCamant, and
  Dawn Song.
\newblock {A Symbolic Execution Framework for JavaScript}.
\newblock In {\em SP}, 2010.

\bibitem{jalangi1}
Koushik Sen, Swaroop Kalasapur, Tasneem Brutch, and Simon Gibbs.
\newblock {Jalangi: A Selective Record-replay and Dynamic Analysis Framework
  for JavaScript}.
\newblock In {\em ESEC/FSE}, 2013.

\bibitem{artform}
Ben Spencer, Michael Benedikt, Anders M{\o}ller, and Franck~van Breugel.
\newblock {ArtForm: A Tool for Exploring the Codebase of Form-based Websites}.
\newblock In {\em ISSTA}, 2017.

\bibitem{wateg}
Suresh Thummalapenta, K.~Vasanta Lakshmi, Saurabh Sinha, Nishant Sinha, and
  Satish Chandra.
\newblock {Guided Test Generation for Web Applications}.
\newblock In {\em ICSE}, 2013.

\bibitem{vogt}
Philipp Vogt, Florian Nentwich, Nenad Jovanovic, Engin Kirda, Christopher
  Kr{\"{u}}gel, and Giovanni Vigna.
\newblock {Cross Site Scripting Prevention with Dynamic Data Tainting and
  Static Analysis}.
\newblock In {\em NDSS}, 2007.

\end{thebibliography}
\end{document}